\documentclass[twocolumn]{aastex631}
\usepackage{amsmath}
\usepackage{pifont}

\shorttitle{Gas metallicity in ram-pressure stripped tails}
\shortauthors{Franchetto et al.}

\begin{document}

\title{Evidence for mixing between ICM and stripped ISM\\
by the analysis of the gas metallicity in the tails of jellyfish galaxies}

\correspondingauthor{Andrea Franchetto}
\email{andrea.franchetto@phd.unipd.it}

\author[0000-0001-9575-331X]{Andrea Franchetto}
\affiliation{Dipartimento di Fisica e Astronomia ``Galileo Galilei'', Universit\`a di Padova, vicolo dell'Osservatorio 3, IT-35122, Padova, Italy}
\affiliation{INAF - Astronomical Observatory of Padova, vicolo dell'Osservatorio 5, IT-35122 Padova, Italy}

\author[0000-0002-8710-9206]{Stephanie Tonnesen}
\affiliation{Flatiron Institute, CCA, 162 5th Avenue, New York, NY 10010, USA}

\author[0000-0001-8751-8360]{Bianca M. Poggianti}
\affiliation{INAF - Astronomical Observatory of Padova, vicolo dell'Osservatorio 5, IT-35122 Padova, Italy}

\author[0000-0003-0980-1499]{Benedetta Vulcani}
\affiliation{INAF - Astronomical Observatory of Padova, vicolo dell'Osservatorio 5, IT-35122 Padova, Italy}

\author[0000-0002-7296-9780]{Marco Gullieuszik}
\affiliation{INAF - Astronomical Observatory of Padova, vicolo dell'Osservatorio 5, IT-35122 Padova, Italy}

\author[0000-0002-1688-482X]{Alessia Moretti}
\affiliation{INAF - Astronomical Observatory of Padova, vicolo dell'Osservatorio 5, IT-35122 Padova, Italy}

\author[0000-0001-5303-6830]{Rory Smith}
\affiliation{Korea Astronomy and Space Science Institute (KASI), 776 Daedeokdae-ro, Yuseong-gu, Daejeon 34055, Republic of Korea}

\author[0000-0003-1581-0092]{Alessandro Ignesti}
\affiliation{INAF - Astronomical Observatory of Padova, vicolo dell'Osservatorio 5, IT-35122 Padova, Italy}

\author[0000-0002-8372-3428]{Cecilia Bacchini}
\affiliation{INAF - Astronomical Observatory of Padova, vicolo dell'Osservatorio 5, IT-35122 Padova, Italy}

\author[0000-0003-3255-3139]{Sean McGee}
\affiliation{School of Physics and Astronomy, University of Birmingham, Birmingham B15 2TT, United Kingdom}

\author[0000-0002-8238-9210]{Neven Tomi\v{c}i\'{c}}
\affiliation{INAF - Astronomical Observatory of Padova, vicolo dell'Osservatorio 5, IT-35122 Padova, Italy}

\author[0000-0003-2589-762X]{Matilde Mingozzi}
\affiliation{Space Telescope Science Institute, 3700 San Martin Drive, Baltimore, MD 21218, USA}

\author[0000-0001-5840-9835]{Anna Wolter}
\affiliation{INAF - Osservatorio Astronomico di Brera, via Brera 28, IT-20121, Milano, Italy}

\author[0000-0001-9184-7845]{Ancla M\"{u}ller}
\affiliation{Ruhr University Bochum, Faculty of Physics and Astronomy, Astronomical Institute, Universit\"atsstr.150, 44801 Bochum, Germany}

%%%%%%%%%%%%%%%%%%%%%%%%%%%%%%%%%%%%%%%%%%

%%%%%%%%%%%%%%%%%%%%%%%%%%%%%%%%%%
%%%%%%% ABSTRACT %%%%%%%%%%%%%%%%%
%%%%%%%%%%%%%%%%%%%%%%%%%%%%%%%%%%
\begin{abstract}

Hydrodynamical simulations show that the ram-pressure stripping in galaxy clusters fosters a strong interaction between stripped interstellar medium (ISM) and the surrounding medium, with the possibility of intracluster medium (ICM) cooling into cold gas clouds. Exploiting the MUSE observation of three jellyfish galaxies from the GAs Stripping Phenomena in galaxies with MUSE (GASP) survey, we explore the gas metallicity of star-forming clumps in their gas tails. We find that the oxygen abundance of the stripped gas decreases as a function of the distance from the parent galaxy disk; the observed metallicity profiles indicate that more than 40\% of the most metal-poor stripped clouds are constituted by cooled ICM, in qualitative agreement with simulations that predict mixing between the metal-rich ISM and the metal-poor ICM.

\end{abstract}

%%%%%%%%%% KEYWORDS %%%%%%%%%%%%%%%%%%
\keywords{\href{http://astrothesaurus.org/uat/584}{Galaxy clusters (584)}; \href{http://astrothesaurus.org/uat/573}{ Galaxies (573)}; \href{http://astrothesaurus.org/uat/2126}{Ram pressure stripped tails (2126)}; \href{http://astrothesaurus.org/uat/1031}{Metallicity (1031)}; \href{http://astrothesaurus.org/uat/832}{Interstellar abundances (832)}; \href{http://astrothesaurus.org/uat/858}{Intracluster medium (858)}}

%%%%%%%%%%%%%%%%%%%%%%%%%%%%%%%%%%%%%%%%
%%%%%%%%%% INTRODUCTION %%%%%%%%%%%%%%%%
%%%%%%%%%%%%%%%%%%%%%%%%%%%%%%%%%%%%%%%%
\section{Introduction}\setcounter{footnote}{8}

Cluster galaxies that move with high velocities ($\sim 1000\,{\rm km\,s^{-1}}$) through the hot and dense intracluster medium (ICM; $T\sim10^7$--$10^8$\,K, $n_e\sim 10^{-4}$--$10^{-2}\,{\rm cm^{-3}}$; \citealt{sarazin1986}) can be affected by ram-pressure stripping (RPS; \citealt{gunn1972}), that is able to eradicate their interstellar medium (ISM).

The interaction between the ISM and the ICM can be extremely complex, and hydrodynamical simulations, that study the mixing process between the two fluids, consider several processes: cloud destruction via Kelvin-Helmholtz instabilities \citep{chandrasekhar1961}
and the heating of the stripped ISM with subsequent evaporation into the ICM \citep{cowie1977a}, as well as the radiative cooling of the ICM onto cold gas clouds \citep{klein1994,gronke2018}.
\citet{tonnesen2021} find that the latter mixing-scenario is possible in a high density and low velocity ICM wind, providing a guide for observational studies.

Observationally, the stripped gas has a multi-phase nature: 
{\sc H\,i} \citep[e.g.][]{chung2009}, molecular gas \citep[e.g.][]{jachym2017}, H$\alpha$ \citep[e.g.][]{gavazzi2001}, and X-ray gas \citep[e.g.][]{sun2010}. In particular, the ionized gas-phase is often due to the photoionization by radiation coming from young stars formed in-situ \citep{poggianti2019}. The optical emission lines of this ionized medium allow us to study several properties of the gas, like the gas-phase metallicity. This quantity can be an excellent tracer to evaluate the mixing scenario between the metal-rich ISM \citep{maiolino2019} and the metal-poor ICM \citep{mernier2018}, bridging the gap between simulations and observations.
Recently, the GASP project (GAs Stripping Phenomena in galaxies with MUSE; \citealt{poggianti2017}) has provided a statistically significant sample of RPS galaxies, probing various gas properties (e.g.\ star formation rate, ionization mechanisms,
kinematics) both in the galaxy disk and in the stripped tails \citep{gullieuszik2017,vulcani2020b,poggianti2019}.
However, neither any of the GASP papers \citep{poggianti2017,gullieuszik2017,bellhouse2019} nor other studies \citep{yoshida2012,fossati2016,merluzzi2016,consolandi2017} focused solely on the gas metallicity in the tails of jellyfish galaxies.\footnote{The term ``jellyfish galaxies'' generally indicates RPS galaxies with spectacular, long gas tails.}

In this Letter we exploit the MUSE observations of the GASP survey to track the gas-phase metallicity along the tails of the most striking jellyfish galaxies to investigate the interaction between the ISM and ICM during RPS. Our analysis provides for the first time key constraints to theoretical models investigating the mixing between the two media.

%%%%%%%%%%%%%%%%%%%%%%%%%%%%%%%%%%
%%%%%%%%%% SAMPLE %%%%%%%%%%%%%%%%
%%%%%%%%%%%%%%%%%%%%%%%%%%%%%%%%%%
\section{Galaxy sample}
To suitably investigate the metallicity profiles in the tails of RPS galaxies, we select from the GASP sample the cluster galaxies with the longest tails and tentacles of ionized gas -- traced by the H$\alpha$ emission -- unilaterally displayed with respect to the main galaxy body. Four galaxies with stripped ionized gas reaching over 50 kpc (in projection) from the galaxy center, represent the best candidates for this study: JO201, JO206, JW100, and JW39.

Since the metallicity profile of the ram-pressure stripped gas from JO201 is already published in \citet{bellhouse2019}, in this Letter we mainly present the results based on the other three galaxies, discussing the implications for JO201 only in the text. The properties of the analyzed galaxies are presented in Tab.~\ref{tab:table1}.

\begin{deluxetable*}{lccccccl}
\tablecaption{Properties of the analyzed galaxies. Columns are: (1) GASP ID number; (2) and (3) Equatorial coordinates of the galaxy center; (4) redshift; (5) logarithm of the stellar mass taken from \citet{vulcani2018b}; (6) gas metallicity at the effective radius from \citet{franchetto2020}; (7) metallicity radial gradient from \citet{franchetto2021}; (8) main references. Further information and images are published in \citet{gullieuszik2020} and on the GASP web site \url{https://web.oapd.inaf.it/gasp/}.}\label{tab:table1}
\tablehead{\colhead{ID} & \colhead{R.A. [J2000]} & \colhead{Decl. [J2000]} & \colhead{z} & \colhead{$\log(M_\star/M_\odot)$} & \colhead{$12+\log({\rm O/H})_{R\rm e}$} & \colhead{$\alpha_{\rm O/H}\,[{\rm dex/R_e}]$} & \colhead{Main reference}\\
\colhead{(1)} & \colhead{(2)} & \colhead{(3)} & \colhead{(4)} & \colhead{(5)} & \colhead{(6)} & \colhead{(7)} & \colhead{(8)}}
\startdata
JO206 & 21:13:47.41 & $+$02:28:34.383 & 0.0489 & $10.96_{-0.05}^{+0.04}$ & $8.96 \pm 0.07$ & $-0.10\pm0.02$ & \citet{poggianti2017}\\
JW100 & 23:36:25.06 & $+$21:09:02.529 & 0.0551 & $11.5_{-0.1}^{+0.1}$ & $9.24 \pm 0.01$ & $-0.04\pm0.01$ & \citet{poggianti2019b}\\
JW39 & 13:04:07.71 & $+$19:12:38.486 & 0.0634 & $11.21_{-0.08}^{+0.07}$ & $9.12 \pm 0.06$ & $-0.04\pm0.01$ & \citet{poggianti2019}\\
\enddata
\end{deluxetable*}

%%%%%%%%%%%%%%%%%%%%%%%%%%%%%%%%%%%%
%%%%%%%%%% ANALYSIS %%%%%%%%%%%%%%%%
%%%%%%%%%%%%%%%%%%%%%%%%%%%%%%%%%%%%
\section{Data analysis}
As extensively explained in \citet{poggianti2017}:
MUSE spectra are corrected for the extinction due to our Galaxy; the stellar-only component is derived applying our spectrophotometric fitting code {\sc sinopsis} \citep{fritz2017} and then subtracted from each spectrum; the gas kinematics, emission line fluxes, and corresponding errors are derived using the software {\sc kubeviz} that fits Gaussian line profiles \citep{fossati2016}; then, the fluxes are corrected for the internal dust extinction
assuming a Balmer decrement of H$\alpha/$H$\beta=2.86$ and applying the extinction law of \citet{cardelli1989}.

We adopt the BPT diagnostic diagrams \citep{baldwin1981} based on the [{\sc N\,ii}]$\lambda6583$ and [{\sc S\,ii}]$\lambda\lambda6717,31$ emission lines to select only star-forming regions whose line ratios are below the separation curves of both \citet{kewley2001} and \citet{kewley2006}. This cross-check more accurately excludes regions powered by ionizing sources different from star formation.

The metallicity of the ionized gas is estimated making use of a modified version of the code {\sc pyqz} (\citealt{dopita2013,vogt2015}; F.~Vogt 2017, private communication). {\sc pyqz} interpolates the observed line ratios {\sc [N\,ii]$\lambda6583$/[S\,ii]$\lambda\lambda6717,31$} and {\sc [O\,iii]$\lambda5007$/[S\,ii]$\lambda\lambda6717,31$} on a model grid computed by {\sc mappings\,iv}, able to disentangle degeneracy with the ionization parameter \citep{sutherland1993,dopita2013}, and delivers the oxygen abundance $12+\log({\rm O/H})$, that here we use as a tracer of the gas-phase metallicity, and the associated error by the propagation of the flux uncertainties.

Disks and tails of jellyfish galaxies are characterized by bright H$\alpha$ clumps, identified to be star-forming clumps surrounded by regions of more diffuse emission \citep{poggianti2019}. These clumps are defined as circular regions centered to the local minima of the Laplace+median filtered H$\alpha$-MUSE image.
The radius of clumps is determined by a recursive algorithm that evaluates the inner counts above the surrounding diffuse emission (see \citealt{poggianti2017} for details).
Spectra within each clump are added and integrated properties are derived as explained before. Clump integrated spectra reach a high signal-to-noise ratio ($(S/N)_{{\rm H}\alpha}\sim90$, on average), therefore, in the following, we focus on these regions. These clumps are sufficiently massive  ($> 10^5 \, M_{\odot}$, \citealt{poggianti2019}) to exclude effects due to incomplete IMF sampling.

The gas-phase metallicity maps of the ionized gas and the identified clumps for JW100, JO206, and JW39 are presented in panel (a) of Figs.~\ref{fig:JW100}, \ref{fig:JO206}, and \ref{fig:JW39}, respectively. The dashed contours in the Figures indicate the galaxy body, defined as the stellar continuum flux 1$\sigma$ above the background level \citep{poggianti2017}.

\subsection{Tails and sub-tails}
The goal of this work is to explore the distribution of the gas-metallicity along the jellyfish galaxy tails. However, the ISM is removed from different positions of the galaxy disk
and the orientation of the stripped gas in the three-dimensional space is unknown.
Therefore, to perform an analysis as accurate as possible, for each galaxy, we identify several main sub-tails, selecting those clumps that are aligned along gas tentacles, and, according to the gas kinematics, have similar line-of-sight velocities. Indeed, observations and simulations show that, in most cases, the stripped gas maintains the rotation it had in the disk \citep{merluzzi2013,gullieuszik2017}. When it is feasible, we also include a maximum of three clumps in the galaxy body to statistically determine the metallicity value of the disk gas in the point where the tail stems from.
Since projection effects prevent a clear distinction, such value is a lower limit because the gas of those clumps might be already stripped.

In panels (b) of Figs.~\ref{fig:JW100}, \ref{fig:JO206}, and \ref{fig:JW39}, we show for each galaxy the clumps belonging to different sub-tails, according to our selection. Focusing separately on each sub-tail allow us to better highlight gas metallicity trends with the projected distance, minimizing the scatter due to the uncertainties on the spatial projection.\footnote{We note that the northern sub-tail of JO206 might be quite uncertain, but removing it from the analysis does not impact on the results.}

To suitably estimate the projected distance of the stripped clumps from the galaxy, we measure their position parallel to the specific stripping direction of the corresponding sub-tail.
This direction is calculated by a linear fit of the clumps of each given sub-tail, and is indicated by an arrow in the panel (b) of the Figures.
Distances are converted in kpc units according to the redshift of the host cluster and then normalized, imposing that the zero-point coincides with the first clump in the sub-tail (including those in the galaxy body). In this sense, these distances have to be interpreted as reference positions measured along the sub-tail, and anyways, these values are lower limits of the real distances, as they are estimated on the sky plane.

We highlight that the results are independent on the choice of the zero-point and, in addition, the metallicity trends persist also when the tail is explored as a whole, but at the expenses of an increased scatter.

%%%%%%%%%%%%%%%%%%%%%%%%%%%%%%%%%%%
%%%%%%%%%% RESULTS %%%%%%%%%%%%%%%%
%%%%%%%%%%%%%%%%%%%%%%%%%%%%%%%%%%%
\section{Results}

\begin{figure*}
    \centering
    \includegraphics[width=0.95\textwidth]{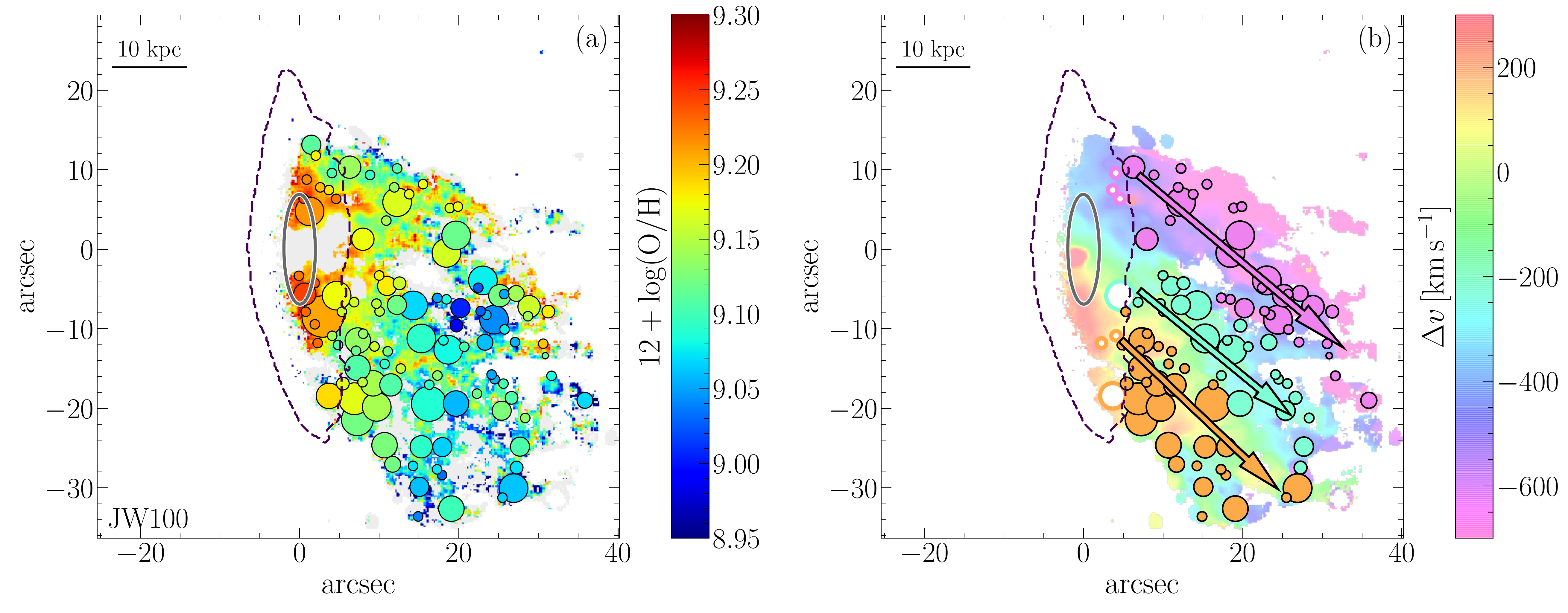}\\
    \vspace{10pt}
    \includegraphics[height=0.3\textwidth]{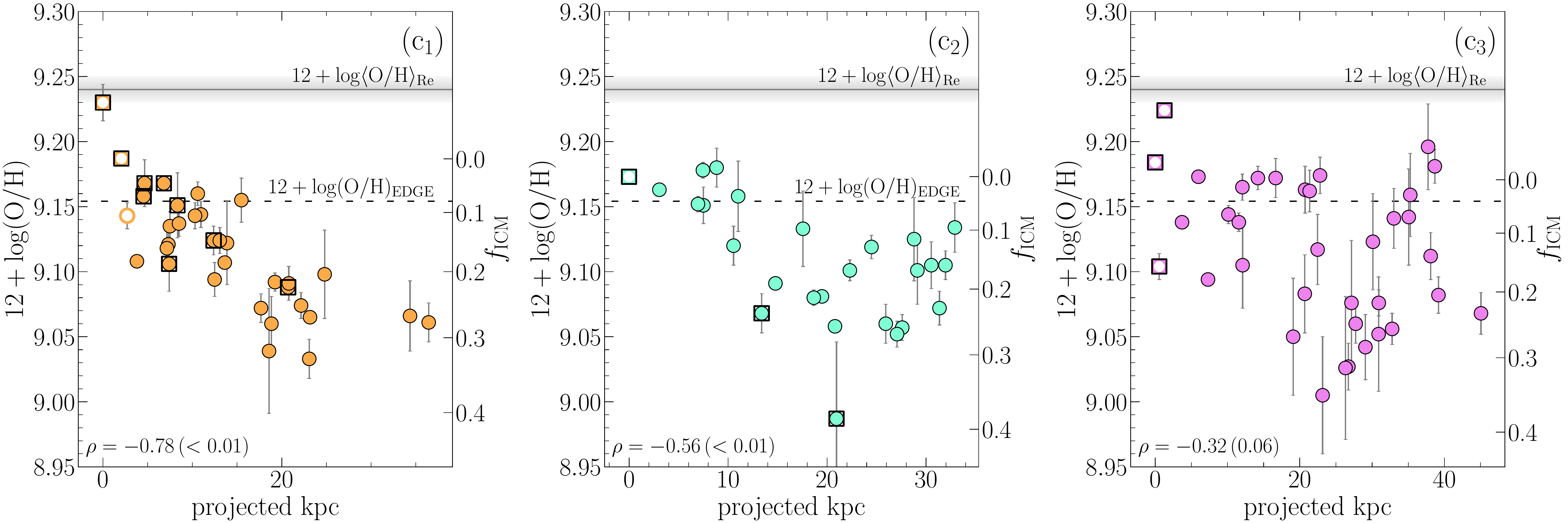}
    \caption{JW100. Panel~(a): gas-phase metallicity map. The circles identify the position and the size of the star-forming clumps. Underlying gray spaxels correspond to the distribution of the H$\alpha$ emission with signal-to-noise ratio greater than 5. Panel~(b): Color-coded gas velocity field map with superimposed clumps belonging to different sub-tails. Each sub-tail is identified by a different color (not correlated to the colorbar). Empty dots correspond to clumps inside the galaxy body. The arrows indicate the direction of the stripping for each sub-tail. In panels~(a) and (b), the dashed contour shows the galaxy body and the gray ellipse indicates the effective radius. Panels~(c): projected metallicity profile of each sub-tail. Dots and errorbars refer to clump metallicities and associated uncertainties, respectively. Square-frames indicate star forming clumps according to [{\sc O\,i}]-BPT. Colors correspond to panel~(b). Gray horizontal lines and the faded areas denote the metallicity at the effective radius and the corresponding uncertainty. Dashed lines indicate the expected metallicity at the edge of the stellar body. In the bottom left corner we also report the Pearson correlation coefficient (and the corresponding p-value) of the observed trends. The scale on the right y-axis indicates the fraction of ICM as explained in the text. \label{fig:JW100}}
\end{figure*}

\begin{figure*}
    \centering
    \includegraphics[width=0.67\textwidth]{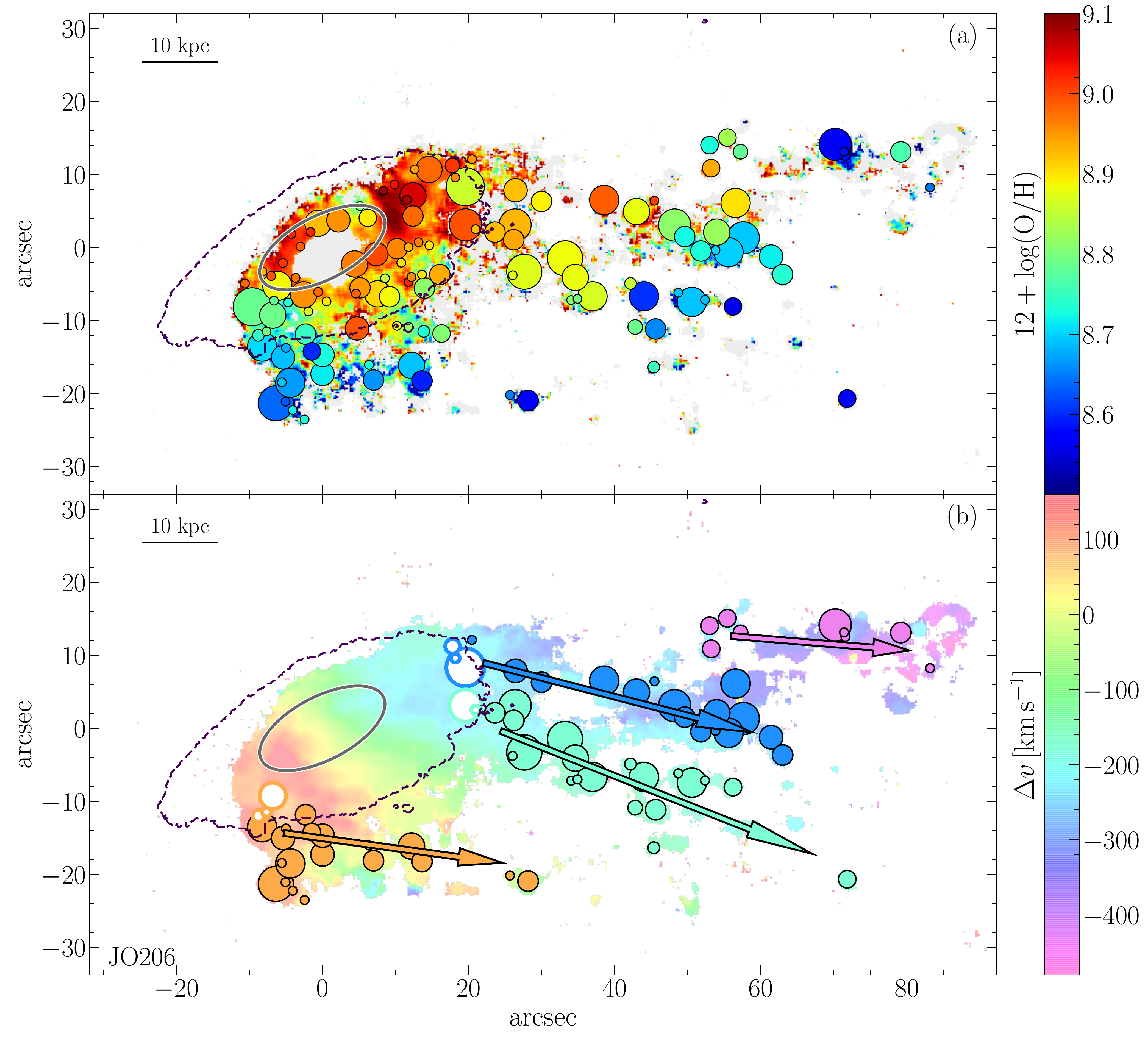}\\
    \vspace{10pt}
    \includegraphics[height=0.6\textwidth]{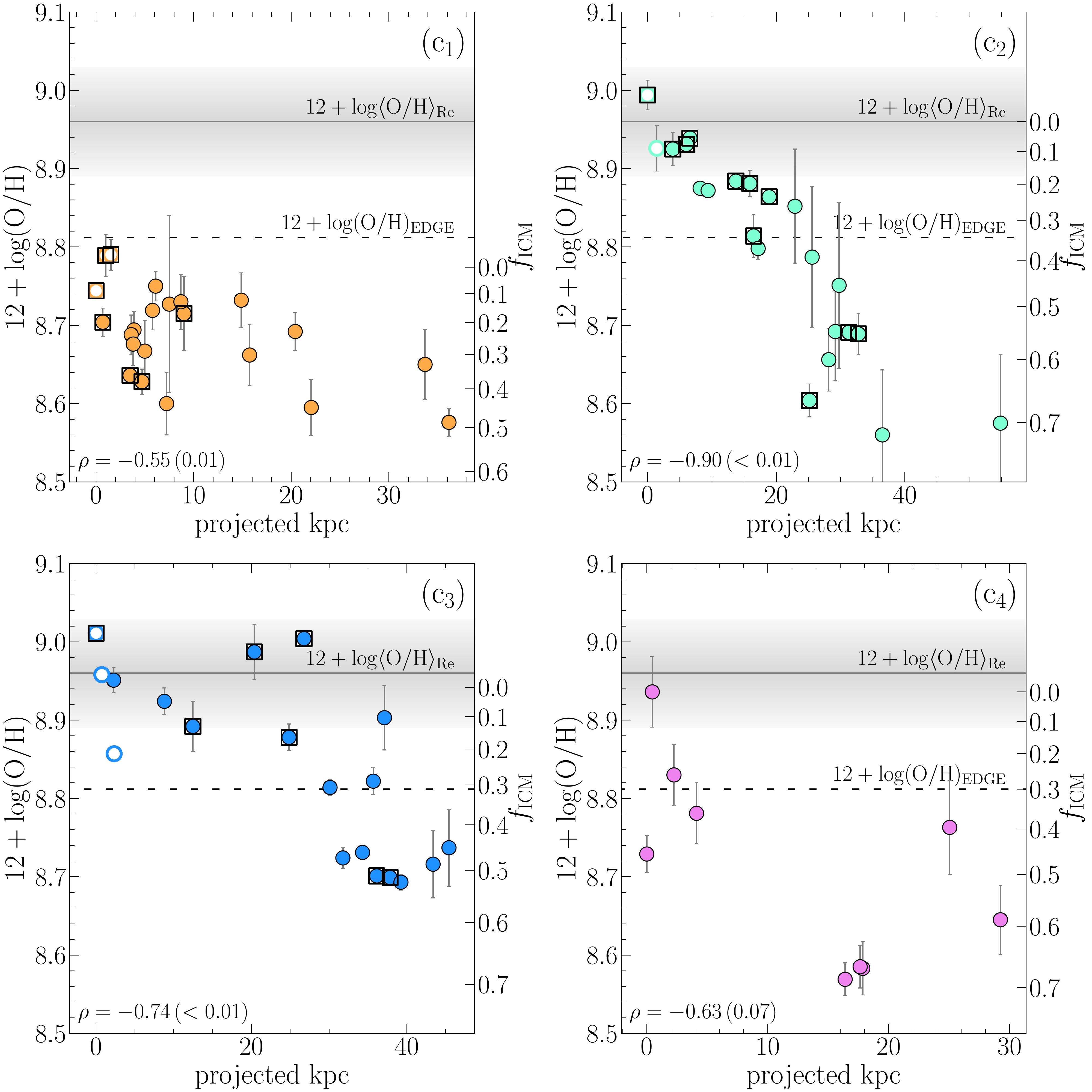}
    \caption{JO206. The panels are as in Fig.~\ref{fig:JW100}\label{fig:JO206}}
\end{figure*}

\begin{figure*}
    \centering
    \includegraphics[width=0.95\textwidth]{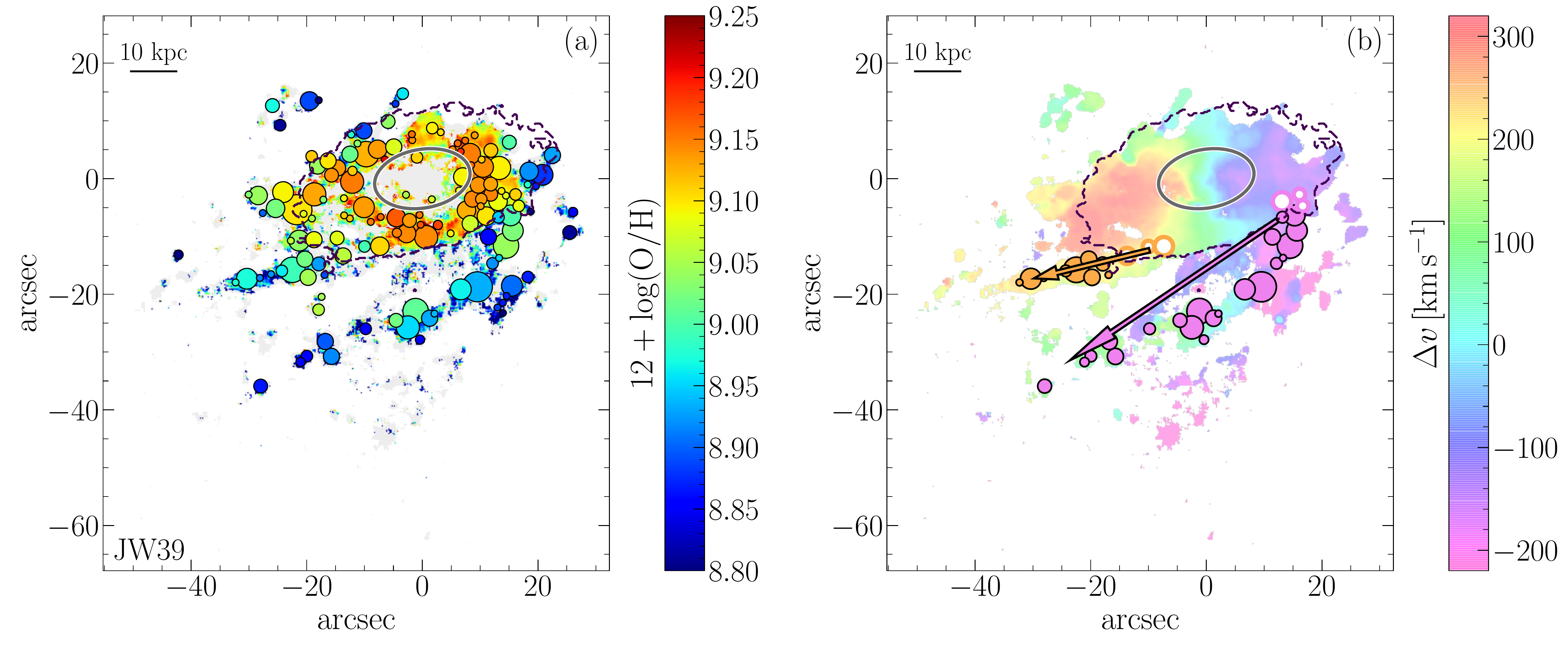}\\
    \vspace{10pt}
    \includegraphics[height=0.3\textwidth]{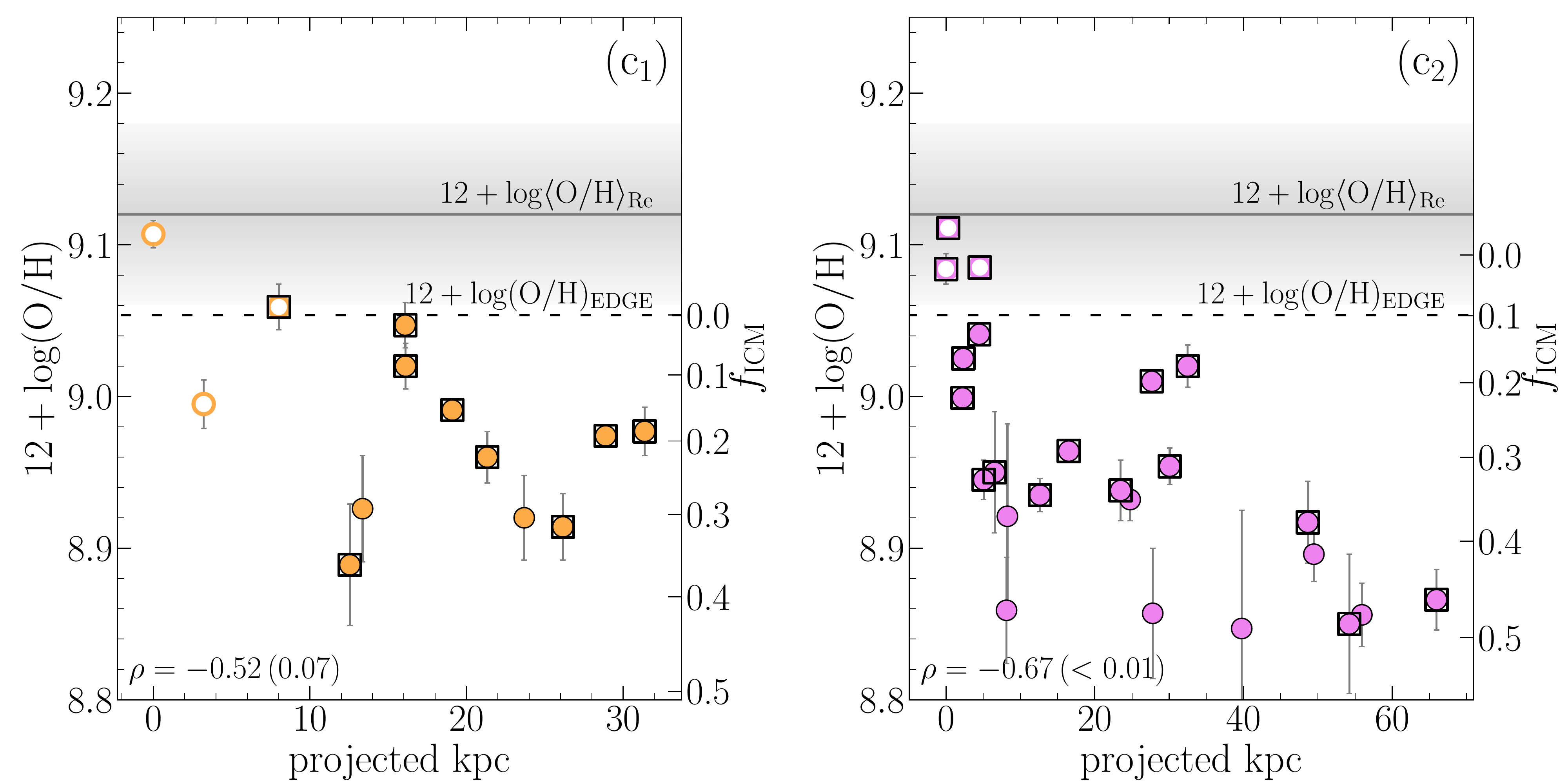}
    \caption{JW39. The panels are as in Fig.~\ref{fig:JW100}\label{fig:JW39}}
\end{figure*}

The panels (a) of Figs.~\ref{fig:JW100}, \ref{fig:JO206}, and \ref{fig:JW39} clearly show that the gas metallicities in the tails of all galaxies are systematically lower than those in the galaxy disks. These trends are better identified looking at the panels (c$_{\rm i}$), where we present the distribution of the metallicities of the clumps in each selected sub-tail 
against their position along the tail.
As a reference of the metallicity in the disk, for each galaxy we indicate (with a horizontal solid line) the value at the effective radius taken from \citet{franchetto2020}. In most cases, metallicities in the tails are much lower than this value. 
Furthermore, in all sub-tails we detect a decrease of the gas metallicity moving along the stripping direction; the decrease varies between $\sim0.1$ and $\sim0.4$~dex, with the lowest metallicities mainly located in the furthest regions. A similar behavior is also observed in JO201, whose tail metallicity trends are very similar to those of the three galaxies presented here; as shown in Fig.~6 of \citet{bellhouse2019}, the metallicities of clumps in the tail decrease by $\sim 0.5$~dex moving away from the galaxy center, along 60 projected kpc of distance. For each sub-tail, we also compute the Pearson correlation coefficient $\rho$ (along with the corresponding p-value) to attest the relevance of these gas metallicity trends; we find a strong decreasing relationship ($|\rho|>0.5$) in most sub-tails with a high statistical significance (p-value $\le 0.07$).

%%%%%%%%%%%%%%%%%%%%%%%%%%%%%%%%%%%%%%
%%%%%%%%%% DISCUSSION %%%%%%%%%%%%%%%%
%%%%%%%%%%%%%%%%%%%%%%%%%%%%%%%%%%%%%%
\section{Discussion and summary}
In previous works, the observed decreasing metallicity profiles along the stripped gas tails have been interpreted as a consequence of the RPS mechanism, which first removes the outermost, metal-poor gas and subsequently the higher metallicity gas closer to the galaxy center, following the metallicity radial profile of galaxies \citep{poggianti2017,gullieuszik2017,bellhouse2019}. Here, we argue that this effect, although plausible, is not sufficient to explain the trends in the analyzed jellyfish galaxies. Indeed, massive galaxies ($M_\star\ge10^{11}\,{\rm M_\odot}$), such as those presented here, have, on average, high gas metallicity values and very shallow metallicity profiles, shallower than intermediate mass galaxies \citep{sanchezmenguiano2016,franchetto2021}. In particular, according to the gas metallicity gradients estimated in the galaxy body of these three galaxies \citep{franchetto2021}, the metallicity values at the edge of the stellar disk (indicated with a dashed line in the panels (c) of the Figs.~\ref{fig:JW100}, \ref{fig:JO206}, and \ref{fig:JW39}) are still much higher than those reached in tails. 

Therefore, we consider three possible scenarios to explain the observed trends and the low metallicity values in the tails:
(1) the progressively lower metallicity gas that is observed further out in the tail might have been removed from the gas disk at large galactocentric radii, beyond the stellar galaxy body, where metallicities are expected to be lower than within the stellar disk;
(2) the particular physical conditions of the gas are leading to an erroneous metallicity estimation; (3) the metallicity values in the tails are due to mixing with the metal-poor ICM. The additional hypothesis that lower density gas, which is more easily stripped, has lower metallicities, is ruled out by our previously published results, showing that gas clumps and diffuse gas have on average similar metallicities \citep{tomicic2021b}.

Scenario (1) would require the tail to narrow with distance and that the gas stripped outside migrates to smaller radii (where the radius is defined as the radius of the approximately cylindrical section of the tail identified by a plane parallel to the galaxy disk at any given distance). 
In particular, assuming the metallicity gradients estimated in the disk of these galaxies \citep{franchetto2021} and hypothesizing that they extend over the stellar body without flattening, the most metal-poor clump gas detected in the tails should have been stripped approximately from galactic radii twice as large as the stellar disk radius.
This is unrealistic for two reasons: simulations show that the tails maintain their radius or, if anything, get wider with distance from the disk \citep{tonnesen2010,tonnesen2021}; both observations and simulations agree that stripped gas maintains its orbital velocity (\citealt{merluzzi2013}; \citealt{gullieuszik2017}; see also top right panels of our Figures), therefore it is not falling to smaller radial distances.

Moreover, looking at our observed tails in their entirety, their radius approximately coincides with the extent of the stellar disk, and all but one of our sub-tails are anchored in the disk. This indicates that the observed ionized gas was not stripped from well beyond the stellar disk.

Scenario (2) considers the possibility that contamination of an additional ionization source 
might alter the observed line ratios. This would yield incorrect metallicity estimates along the tail because the model grids adopted by {\sc pyqz} are designed including only photoionization from young massive stars. 
Evidence for additional ionization mechanisms in some tails, likely due to the interaction ICM-ISM, is based on an excess of the [{\sc O\,i}]$\lambda6300$ line \citep{poggianti2019,campitiello2021,tomicic2021b}.
Although this effect is seen especially in the diffuse gas (outside of clumps) and we spent particular attention to select only star-forming clumps, we perform a further check excluding those clumps that are not powered by star-formation according to the BPT diagram based on the [{\sc O\,i}] line, adopting the separation curve of \citet{kewley2006}. Although this selection preserves only a third of the valid clumps (square-framed dots in bottom panels of Figs.~\ref{fig:JW100}, \ref{fig:JO206}, and \ref{fig:JW39}), in all three galaxies we still find a clear metallicity profile along the tails.
As additional check, we also re-analyze the profiles computing the metallicities with the O3N2 calibration of \citet{curti2017} (plots not shown) finding that the correlation still holds. This confirms that the observed trends are real and not due to a systematic bias in the metallicity estimates.
However, in principle, we cannot rule out the effect of exotic processes that might produce artifacts in the metallicity measurement, but we are not in the position to evaluate it.

Scenario (3) invokes the mixing between the stripped ISM and the ICM, observed in some simulation works \citep[e.g.][]{gronke2018,tonnesen2021}.
X-ray observations find that clusters are characterized by a uniform iron abundance of
$Z_{\rm Fe} \sim 0.3$ Solar beyond 0.2--0.3 virial radius \citep{mernier2018}. As our galaxies present super-solar metallicities in their disks, a mixture of these two components could produce an evident decrease of metallicity proportional to the amount of ICM cooled in the stripped material. A rough estimate of the ICM fraction in the clumps can be done assuming that the observed metallicity in tails is a linear combination of the stripped ISM and ICM metallicities weighted by the fraction of the two components:
\begin{equation}
    Z_{\rm obs}=Z_{\rm ISM}\,f_{\rm ISM}+Z_{\rm ICM}\,f_{\rm ICM},
\end{equation}
where the metallicities are expressed in solar units\footnote{$\log(Z[{\rm Z_\odot}])=12+\log({\rm O/H})-8.69$, where 8.69 is the solar oxygen abundance adopted by {\sc pyqz}.}
and $f_{\rm ISM}+f_{\rm ICM}=1$.

Unfortunately, $Z_{\rm ISM}$ and $Z_{\rm ICM}$ are not well constrained, therefore we need to make some assumptions, trying to be as conservative as possible. In absence of an accurate measure of the ICM metallicity around our galaxies, we set $Z_{\rm ICM}=0.3$.
The $Z_{\rm ISM}$ of each sub-tail is the mean metallicity of corresponding clumps inside the disk, or in absence of them the highest value observed among the selected clumps.
By this simple calculation, we can obtain the $f_{\rm ICM}$ values corresponding to metallicities detected in each sub-tail. These values are reported on the right y-axis of the panels (c) of the Figs.~\ref{fig:JW100}, \ref{fig:JO206}, and \ref{fig:JW39}. For JW100 and JW39 we derive that $\sim 30$--$40\%$ of the most metal-poor clumps could be constituted by gas from the ICM; for JO206 the mixing can reach values as high as $\sim 60\%$; while, for JO201 the effect is evaluated up to the 80\%.

We stress that these values are only indicative as they can suffer by large uncertainties due to the simplistic estimation. 
Despite these qualifications, we can make a qualitative comparison with the simulated RPS galaxies presented in \citet{tonnesen2021}, who study the mixing of the galactic gas with the ICM wind.
In detail, we select similar regions from the tail of one of their three RPS galaxies (HDLV from that work).
We use the clump finder routine in {\sc yt} \citep{turk2011} to find groups of at least 200 connected cells above the threshold density of 10$^{-26}\,{\rm g\,cm^{-3}}$, 
for which we compute their mass-weighted positions; to mimic the MUSE observations, we select spherical 1 kpc regions centered on these positions with at least 20 dense cells ($\rho > 10^{-24}\,{\rm g\,cm^{-3}}$); these dense cells are used to derive the mass-weighted ICM fraction of each of these clumps and their distance from the galaxy disk.

\begin{figure}[t]
    \centering
    \includegraphics[width=\columnwidth]{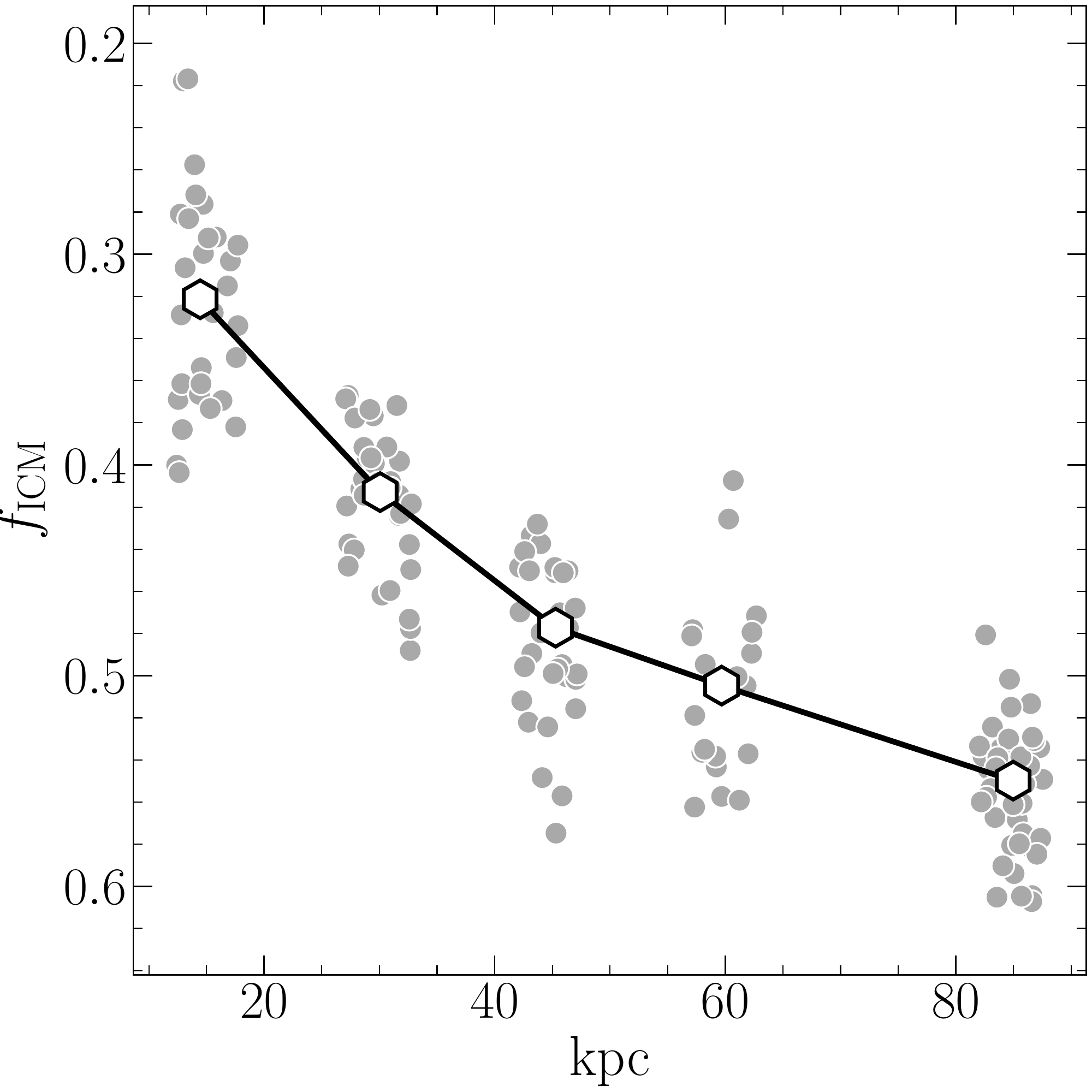}
    \caption{The distribution of $f_{\rm ICM}$ in clumps identified in a snapshot of the tail of a simulated RPS galaxy (440 Myr in HDLV from \citealt{tonnesen2021}) as a function of the distance from the galaxy. Gray dots refer to the values of clumps, while white hexagons indicate the median value at each given distance.}
    \label{fig:simulation}
\end{figure}

In Fig.~\ref{fig:simulation}, we find a clear trend between the $f_{\rm ICM}$ in clumps and their physical distance from the simulated galaxy: moving from 15~kpc to 85~kpc, the amount of ICM cooled in the clumps increases from 32\% to 55\%, on average, of the total clump mass, with fluctuations inside a spread of 10\%. While the exact values vary, the trend holds in all three \citet{tonnesen2021} simulations across hundreds of Myrs, and is discussed in detail in \citet{tonnesen2021}. Although a direct quantitative comparison is not possible due to the differences in the ICM properties, galaxy masses, and orbits, as well as projection effects, simulations agree with our observations and support the ISM-ICM mixing scenario.

Finally, since we are focusing on gas of clumps involved in recent star formation, we should expect a metal enrichment due to the stellar yields, conversely to observations. Therefore, finding low metallicities contributes to reinforcing the hypothesis of mixing.

To summarize, in this paper we showed three cases of ram pressure stripped galaxies where star-forming clumps in the tails have lower metallicity at larger galactic distances. The metallicity decrease is much larger than the expected trend due to the outside-in stripping, and we argue that ISM-ICM mixing is needed, in good agreement with predictions from simulations. 
Thus, a picture in which clouds are stripped intact directly from the disk and survive unmixed is inconsistent with the observations shown here. 
Even dense gas clumps seem to be well-mixed with the ICM.

%%%%%%%%%%%%%%%%%%%%%%%%%%%%%%%%%%%%%%%%%%%%%%%%%%%%%%%%%%%
%%%% ACKNOWLEDGMENTS %%%%%%%%%%%%%%%%%%%%%%%%%%%%%%%%%%%%%%
%%%%%%%%%%%%%%%%%%%%%%%%%%%%%%%%%%%%%%%%%%%%%%%%%%%%%%%%%%%
\vskip 5mm
\noindent
We thank the GASP team for useful discussions.
Based on observations collected at the European Organization for Astronomical Research in the Southern Hemisphere under ESO programme 196.B-0578. This project has received funding from the European Research Council (ERC) under the European Union's Horizon 2020 research and innovation programme (grant agreement No.\ 833824, PI Poggianti). We acknowledge funding from the INAF main-stream funding programme (PI B.~Vulcani). B.~V.\ and M.~G.\ acknowledge the Italian PRIN-Miur 2017 (PI A.~Cimatti). S.~M.\ acknowledges support from the Science and Technology Facilities Council through grant number ST/N021702/1.
%\begin{acknowledgements}
%\end{acknowledgements}

%%%%%%%%%%%%%%%%%%%%%%%%%%%%%%%%%%%%%%%%%%%%%
%%% SOFTWARE %%%%%%%%%%%%%%%%%%%%%%%%%%%%%%%%
%%%%%%%%%%%%%%%%%%%%%%%%%%%%%%%%%%%%%%%%%%%%%
\software{{\sc sinopsis} \citep{fritz2017}, {\sc kubeviz} \citep{fossati2016}, {\sc pyqz} \citep{dopita2013,vogt2015}, {\sc yt} \citep{turk2011}}

%%%%%%%%%%%%%%%%%%%%%%%%%%%%%%%%%%%%%%%%
%%%% BIBLIOGRAPHY %%%%%%%%%%%%%%%%%%%%%%
%%%%%%%%%%%%%%%%%%%%%%%%%%%%%%%%%%%%%%%%
%\bibliography{myref}

\end{document}